\documentclass[12pt,twoside]{article}
\input BoxedEPS
\SetRokickiEPSFSpecial  
\HideDisplacementBoxes

\begin{document}
\pagestyle{myheadings}

\begin{center}
{\Large {\bf 
Superscaling of Inclusive Electron Scattering\\
from Nuclei} }
\end{center}

\markboth{Superscaling of Inclusive Electron Scattering 
from Nuclei}{T.W. Donnelly and Ingo Sick}

\bigskip

\thispagestyle{empty}

\begin{center}
T. W. Donnelly \\
Center for Theoretical Physics,  Laboratory for Nuclear Science \\
and Department of Physics \\
Massachusetts Institute of Technology \\
Cambridge, Massachusetts 02139-4307, USA  \\[5mm] 
Ingo Sick \\
 Departement  f\"ur Physik und Astronomie,
Universit\"at Basel  \\ CH4056 Basel, Switzerland 
\end{center}

\bigskip

\begin{center}
\begin{minipage}{12.cm}
\small
We investigate the degree to which the concept of superscaling, initially 
developed within
the framework of the relativistic Fermi gas model, applies to inclusive 
electron scattering from nuclei. We find that data obtained from the low 
energy loss side of the quasielastic peak exhibit the superscaling property, 
{\em i.e.} the scaling functions $f(\psi ')$  are not only 
independent of momentum transfer (the usual type of scaling: scaling of the 
first kind), but coincide for $A \geq 4$ when plotted versus 
a dimensionless scaling variable $\psi '$ (scaling of the 
second kind). We use this behavior to study as yet poorly understood
properties of the inclusive response at large electron energy loss.
\end{minipage}

\end{center}
\normalsize

\section{Introduction \label{int}}

The applications of scaling and dimensional analysis have been
important tools for the development of new insights in physics. Scaling 
in scattering experiments is
observed in processes where a weakly interacting probe scatters from 
constituents bound in a composite system and a constituent is ejected 
quasifreely from the system. The (unpolarized) inclusive response functions, 
determined by observing only the scattered probe, in addition to depending on 
the scattering angle, in general depend explicitly on only two more 
independent 
variables --- the energy  $\omega$ and momentum  {\bf q} transferred by the 
probe to the constituent.  In the asymptotic regime of large $q=|{\bf q}|$ 
and $\omega$,
however, when appropriately divided by the elementary probe-constituent cross 
section, the responses are (approximately) functions of only 
a {\em single} variable $z = z(q,\omega)$, with $z$ in turn  a function 
of $q$ and $\omega$. This functional independence of the so-called scaling 
function on the momentum transfer (which sets the scale in the scattering) 
is known as scaling and is seen as a signature that the scattering occurred 
between the probe and the specific elementary constituent of the target,
rather than arising from some other process such as scattering from 
different constituents. To distinguish this behavior
from the additional scaling that forms the focus of the present work
we call the usual independence of momentum transfer {\em scaling of the 
first kind}. Various choices for the function $z(q,\omega)$ can be 
motivated on the basis of the kinematics of the probe-constituent elastic 
scattering process --- several such choices are discussed in this work. 
Expressed in terms of a  scaling function, the inclusive cross sections can 
be related to the momentum distribution (more generally, to the spectral
function) of the constituents in the target.

In the last 20 years or so, the concept of $y$-scaling in scattering of
high-energy electrons from nuclei has  been actively pursued \cite{Day90}. 
For $y$-scaling the focus is on protons and neutrons in nuclei as the 
``elementary'' constituents. Typically, when three-momenta of $q > $500 MeV/c 
and energies at or somewhat below the quasielastic peak position
$\omega \approx (q^2+m_N^2)^{1/2}-m_N$, where $m_N$ is the nucleon mass, 
are transferred from the electron to the 
nucleus via exchange of a virtual photon, a nucleon is ejected from the nucleus
in a reasonably ``quasifree'' manner. Namely, the nucleon 
leaves the nucleus with a high enough energy that the process can be treated 
approximately as having occurred without strong effects from 
final-state interactions (FSI).
Under the appropriate kinematical conditions (to which we return below)
the cross section can be written as a product of the elementary 
electron-nucleon elastic cross section times a function $F$. It has been 
shown theoretically and verified experimentally that at large momentum 
transfers the appropriately defined function $F$  depends only on a single 
variable $y=y(q,\omega)$, itself  a function of $q$ and $\omega$; here $y$
 is a particular choice for the general function $z$ referred to above
(see also the next section). The scaling 
function so obtained asymptotically contains interesting information about 
the dynamical properties of the nuclear ground state, and the fact that 
scaling does occur provides very useful information about the reaction 
mechanism itself.  

Indeed, scaling in electron-nucleus scattering is a special case of a more 
general phenomenon \cite{West75} occurring in various areas of physics
that deal with inelastic scattering of a weakly interacting probe from a 
many-body system in which $q$ and $\omega$ are transferred to a single 
constituent in the target system. Examples are found in the scattering of keV 
electrons from electrons bound in atoms \cite{Bonham77}, in the scattering of 
eV neutrons from atoms in solids or liquids \cite{Hohenberg66} and in 
the scattering of GeV electrons from quarks in the nucleon \cite{Friedman72}. 
Despite the extraordinary range of energy and momentum transfers for which 
these reactions have been studied, the conceptual ideas used to describe the 
scaling phenomena in these different  fields have many features in common.

For electron-nucleus scattering, the topic of particular interest to this 
paper, scaling was already implicit in the early theoretical studies of
electron-nucleus 
quasielastic scattering \cite{Czyz63} when treated in terms of the 
non-relativistic Fermi gas model; the cross section could be reduced to a 
function of a single variable multiplied by the elementary e-N elastic 
cross section. For treating electron scattering explicitly in 
terms of $y$-scaling, the seminal idea originated with the work of 
West \cite{West75}. Early theoretical work was also undertaken by 
Kawazoe \cite{Kawazoe75}. At that time, however, few data were available and 
where they did exist they were restricted to a narrow kinematical 
range at low energy. Conclusive observation of the asymptotic 
$q$-independence of $F$ for inclusive scattering became possible with the 
availability of data that spanned a large range of 
$q$ and $\omega$ and was presented by 
Sick {\em et al.} \cite{Sick80} for $^3$He. Subsequent work 
\cite{Pace82,Ciofi83} placed the theoretical foundations of scaling on 
reasonably firm ground, specifically addressing a variety of issues such as 
the role of the restrictions imposed by the nature of the 
$(A-1)$ system excitation spectrum and recoil-nucleon FSI 
\cite{Benhar91}. A summary of the various aspects of conventional
(first kind) scaling has been 
given in the review paper of Day {\em et al.} \cite{Day90}. 

Much of the previous work has concentrated on the study of the scaling 
properties of the response in the low-$\omega$ tail of the quasielastic peak; 
in this region the scaling function is sensitive to components of the 
spectral function at large initial nucleon momenta. Particular emphasis was 
placed on light nuclei $A\leq$4 where the scaling approach works particularly 
well and where sophisticated calculations of ground-state nuclear wave
functions and hence spectral functions are available for theoretical studies 
of the scaling properties.

In the present paper, we explore a different aspect of scaling. Rather than
concentrating on the response of individual nuclei, we compare the 
scaling function of {\em different nuclei} with $A\geq$4, and study the degree
to which these scaling functions are the {\em same} --- we call such
behavior scaling of the {\em second kind}. The motivation is to explore
the degree to which the concept of {\em superscaling} introduced by 
Alberico {\em et al.}  \cite{Alberico88} when studying the properties of the 
Relativistic Fermi Gas (RFG) model, that is, scaling of
both the first and second kinds, is applicable to nuclei. A 
presentation in condensed form of some of this analysis is available 
in Ref.~\cite{Donnelly99}.

Here we study superscaling using a large body of 
inclusive scattering data. While we employ  the RFG model to motivate the 
choice of the scaling variable, only minimal use of this model is 
subsequently made in interpreting the data as the actual dynamical physics 
content in the problem is undoubtedly more complex than the RFG model can be
expected to address. The RFG does, however, offer a physical scale 
--- the Fermi momentum --- that can be used to make both the scaling variable 
and scaling function dimensionless. The emphasis of the present paper therefore 
is on superscaling as observed in the experimental data, and on the 
physics one can deduce from this scaling property. Here (in 
Sec.~\ref{sec:formal1} and the Appendix) we also provide
discussion in depth of the choices made for the scaling variables,
their limiting expressions and the inter-relationships amongst them.
  
Scaling of the second kind and superscaling were actually implicit 
although unrecognized in the early work of 
\cite{Whitney74} in which quasielastic scattering was studied at one value
of momentum transfer for a range of nuclei $A =$ 6--208. The data at 
energy loss below the maximum of the quasielastic peak could be explained in
the Fermi gas model, an observation that implies that scaling of the 
second kind did occur. No scaling analysis of the data was performed,
however.

Using modern data we find that the superscaling idea works very well in the 
region below the quasielastic peak, as discussed in Sec.~\ref{sec:results1} 
(see also \cite{Donnelly99}). However, some breaking of superscaling does 
occur, and in the present work we use such deviations to elucidate some of the 
as yet not well understood features observed in the various
measurements of the 
quasielastic response. In particular, in Sec.~\ref{sec:lt} we focus on the 
difference between the longitudinal and transverse responses in the region 
of the quasielastic peak, and the properties of the contributions that fill 
in the ``dip'' between the quasielastic and $\Delta$ peaks --- a region which
has presented a puzzle for a long time \cite{Whitney74}.     
 
Following this introduction we proceed in Sec.~\ref{sec:formal1} to 
discuss the relevant formalism involved in scaling  
of the first and second kinds for the cross sections, relegating some
details to an Appendix.  Then in Sec.~\ref{sec:results1} we discuss the 
results of analyzing  the existing data on quasielastic electron 
scattering, including recent results from TJNAF \cite{Arrington99}, to test the 
idea of superscaling. Subsequently in 
Sec.~\ref{sec:lt} we  specialize the formalism and discussion of the data to a 
treatment of the individual longitudinal and transverse responses. We end 
in Sec.~\ref{sec:concl} with the conclusions to be drawn from the present 
study and with some discussion of the questions that must still be regarded 
as open ones in studies of quasielastic electron scattering from nuclei at 
intermediate energies.

\section{ Scaling of Cross Sections: Formalism }
\label{sec:formal1}
Let us begin by repeating and extending some of the arguments that underlie
the concept of $y$-scaling of the unseparated cross sections; 
these form part of the basis of the 
discussions of superscaling that follow, and  several identities
and relationships amongst the variables involved are presented for the
first time. In the usual approach to inclusive electron scattering in the 
quasielastic regime one assumes that the dominant process is impulsive 
one-body knockout of nucleons together with contributions from two-body 
processes that play a role when the normally dominant process is suppressed. 
Of course, for the ideas of scaling to be applicable one must avoid the
regime of low energy and momentum transfers where strong FSI 
effects (including Pauli blocking, collective behavior in
the final states, etc.) are felt. Also the distortion of the
initial and final electron wave functions moving in the Coulomb field
of the nucleus must be addressed (at least for heavy nuclei and low electron 
energy) for the
scaling analysis to be clear.

The usual approach is the following: One starts from the 
$(e,e'p)$ and $(e,e'n)$ cross sections which may be written as functions of the
3-momentum transfer $q=|{\bf q}|$ and energy transfer $\omega$, 
the electron scattering angle $\theta_e$, the azimuthal 
angle $\phi_N$ between the planes in which the electrons lie and 
in which the momentum transfer and the outgoing nucleon 
lie, and two variables specifying the remaining kinematics of the outgoing 
nucleon. For the latter one may use the 3-momentum of the nucleon 
$p_N=|{\bf p}_N|$ or its energy $E_N=(m_N^2+p_N^2)^{1/2}$ and its polar
angle $\theta_N$, the angle between {\bf q} and ${\bf p}_N$. Alternatively
one may use the magnitude of the missing-momentum $p=|{\bf p}|=
|{\bf p}_N-{\bf q}|$ and a variable to characterize the degree of
excitation of the residual system; for the latter we use 
\begin{equation}
{\cal E}(p) \equiv \sqrt{(M_{A-1})^2 + p^2} - \sqrt{(M_{A-1}^0)^2 + p^2}
  \geq 0,
\label{scriptE}
\end{equation}
where $m_N$ is the nucleon mass, $M_{A-1}$ is the (in general) excited 
recoiling system's mass, while $M_{A-1}^0$ is that system's mass when
in its ground state. The target mass is denoted $M_A^0$ and the separation
energy relates three of the masses in the following way: 
$E_S\equiv M_{A-1}^0+m_N-M_A^0 \geq 0$. The variable ${\cal E}$ is essentially
the familiar missing-energy minus the separation energy.

The strategy in the usual scaling analyses is to determine the smallest
value of missing-momentum $p$ that can occur for the smallest possible
value of missing-energy (i.e., ${\cal E}=0$), since there, at least for 
kinematics not too far removed from the quasielastic peak, one might expect
to have the largest contributions from the underlying nuclear spectral
function --- see, however, the discussions at the end of this section. 
This smallest value of $p$ is traditionally defined to be
$y$ ($-y$) for $\omega$ larger (smaller) than its value at the quasielastic
peak. Thus one may use $(q,y)$ rather than $(q,\omega)$ as the two variables
together with $\theta_e$ upon which the inclusive cross section depends.
In the Appendix we give complete expressions for $y$ and for the largest
value of $p$ that may be reached for given $q$ and $\omega$; here we give
an expression only for $y$ in the limit where $M_{A-1}^0 \to \infty$, as 
for all but the very lightest nuclei this is an excellent approximation:
\begin{equation}
y_\infty=\sqrt{{\widetilde\omega}(2m_N + {\widetilde\omega})} - q,
\label{yinfty}
\end{equation}
where ${\widetilde\omega}\equiv \omega - E_S$. Corrections of
order $(M_{A-1}^0)^{-1}$ are also given in the Appendix (see 
Eq.~(\ref{yexpand})).

Focusing now on the region $y<0$, the most common approach to $y$-scaling 
(see, for example, \cite{Day90}) is then to evaluate the
single-nucleon cross section at the lowest values of 
$(p,{\cal E})$ that can be reached for given values of $q$ and $y$ --- 
in the scaling region these are $p=-y$ and ${\cal E}=0$ --- and then to
divide the inclusive cross section by this quantity to define a function
of $q$ and $y$: 
\begin{equation}
  F(q,y) \equiv \frac{ d^2\sigma / d\Omega_e d\omega }
    { {\widetilde\sigma}_{eN}(q,y;p=-y,{\cal E}=0) } .
\label{Fscaling}
\end{equation}
For the single-nucleon cross section
it is common practice to use the cc1 prescription of De Forest \cite{Forest83}
with the form factors parametrized as in \cite{Hoehler76}.
In the Appendix we provide the complete expressions for the cross section and
indicate the degree to which the struck nucleon in Plane-Wave
Impulse Approximation (PWIA) is off-shell. In PWIA one has
\begin{equation}
  F(q,y) = 2\pi \int\limits^Y_{-y} p \, dp\, \widetilde{n} (q, y; p) ,
\label{Fintn}
\end {equation}
involving the integral
\begin{equation}
  \widetilde{n} (q, y; p) = \int\limits^{{\cal E}_M}_0 d{\cal E} 
    \widetilde{S}(p,{\cal E}) 
\label{nints}
\end{equation}
whose upper limit is approximately given by 
\begin{eqnarray}
{\cal E}_M (q,y;p) &\cong& \sqrt{m_N^2+(q+y)^2}-\sqrt{m_N^2+(q-p)^2} 
  \label{maxmissappx1} \\
&\cong& m_N + {\widetilde\omega} -\sqrt{m_N^2+(q-p)^2} 
\label{maxmissappx2}
\end{eqnarray}
using Eq.~(\ref{yinfty}). Here again we have taken the limit where 
$M_{A-1}^0 \to \infty$, relegating
the exact expressions to the Appendix (see Eqs.~(\ref{maxmiss})).
At high enough values of $q$ one seeks the $y$-scaling behavior: namely,
if the inclusive response scales then $F$ becomes only a function of $y$,
\begin{equation}
F(q,y) \stackrel{q \to \infty}\longrightarrow F(y)\equiv F(\infty,y) .
\label{scaling}
\end{equation}

Scaling has also been approached from a different point of view using
as a starting point the RFG model \cite{Alberico88,Barbaro98}.
The strategy there is to provide a form similar to Eq.~(\ref{Fscaling})
such that in this model exact scaling is obtained. As seen in 
\cite{Barbaro98} the variable
\begin{equation}
y_{RFG} = m_N \left( \lambda \sqrt{1+1/\tau} - \kappa \right)
\label{yyrfg}
\end{equation}
naturally emerges. Here, as in many past studies, we employ dimensionless 
versions of $q$, $\omega$ and $|Q^2|$: $\kappa\equiv q/2 m_N$,
$\lambda\equiv \omega/2 m_N$ and $\tau\equiv |Q^2|/4 m_N^2=\kappa^2
-\lambda^2$. Below we show how to inter-relate the variables $y$ and 
$y_{RFG}$. As also discussed in the above-cited work, 
a dimensionless scaling variable $\psi$ is strongly motivated by the
RFG model. In the Appendix we give the exact expression for $\psi$ (see
Eq.~(\ref{psifull})), whereas
for most purposes the following approximations are excellent:
\begin{eqnarray}
  \psi &=& \frac{y_{RFG}}{k_F} 
   \biggl[1+{\cal O} [\eta_F^2] \biggr]  \label{psi}\\
   &\cong& \frac{1}{\eta_F} \biggl[ \lambda \sqrt{1+1/\tau}
   -\kappa \biggr] ,
\label{psiappxgood}
\end {eqnarray}
where $k_F$ is the Fermi momentum and $\eta_F = k_F/m_N$ its dimensionless
counterpart. Typically $\eta_F$ is small, growing from 0.06 for deuterium 
to about 0.3 for the heaviest nuclei, and thus expansions such as those above
are usually quite good, since they neglect terms only of order
$\eta_F^2$. An alternative approximation for $\psi$ that also 
proves useful to introduce is the following:
\begin{equation}
   \psi = \psi_0 \biggl[ 1 + \sqrt{1+1/4\kappa^2} \frac{1}{2} \eta_F
   \psi_0 + {\cal O} [\eta_F^2] \biggr] ,
\label{psiappx}
\end{equation}
where now the good (but not as good) variable
\begin{equation}
\psi_0 \equiv \frac{2}{\eta_F} \biggl[ \sqrt{\lambda (1+\lambda)} 
  -\kappa \biggr] 
\label{psi0}
\end{equation}
occurs and the result in Eq.~(\ref{psiappx}) receives linear
(rather than quadratic) corrections when written in terms of $\psi_0$.

The RFG analog of Eq.~(\ref{Fscaling}) is
\begin{equation}
F (\kappa, \psi) \cong \frac{d^2 \sigma/d \Omega_e d \omega}
   {\sigma_M [\frac{\kappa}{2\tau} v_L {\widetilde G}^2_E+ 
   \frac{\tau}{\kappa} v_T {\widetilde G}^2_M]} ,
\label{Ftotal}
\end{equation}
where we have made use of 
the usual lepton kinematical factors $v_L$ and $v_T$ and the approximations
for the single-nucleon responses $G_L$ and $G_T$ which involve
${\widetilde G}^2_E\equiv ZG^2_{Ep}+NG^2_{En}$ with ${\widetilde G}^2_M$
defined similarly (see Eqs.~(\ref{formfac}) and
(\ref{Ftotalfull}--\ref{Delta})). Note that relativistic factors
involving the difference between $\kappa^2$ and $\tau$ 
in Eq.~(\ref{Ftotal}) are very important
to retain when studying quasielastic scattering at high momentum transfers.

Before carrying these ideas over to an analysis of the data, it is useful
to bridge the gap between the usual $y$-scaling approach and the
$\psi$-scaling ideas contained in the RFG (see also \cite{Barbaro98}).
First, let us use the $y$ variable to define its dimensionless counterpart,
\begin{equation}
\Upsilon \equiv y/k_F
\label{ups}
\end{equation}
and from Eq.~(\ref{yinfty}) its approximate form,
\begin{equation}
\Upsilon_\infty \equiv y_\infty /k_F 
  = \frac{2}{\eta_F} \biggl[ \sqrt{ {\widetilde\lambda} (1+ 
  {\widetilde\lambda}) } - \kappa \biggr] ,
\label{upsinfty}
\end{equation}
where ${\widetilde\lambda} \equiv {\widetilde\omega}/2 m_N$. Clearly the 
two approaches will yield rather similar results, since
\begin{equation}
\psi_0 = \Upsilon_\infty (E_s =0) ,
\label{leading}
\end{equation}
with corrections to $\Upsilon$ coming from the finite-mass effects 
discussed in the Appendix (see Eq.~(\ref{yexpand})) and to $\psi$ 
from the $\eta_F$-dependent terms in Eq.~(\ref{psiappx}) (see also below). 
The $y$- and $\Upsilon$-variables build in the kinematics of
nucleon knockout and recognize the initial-state separation energy $E_s$;
however, they do not take into account the missing-energy dependence in
the cross section. On the other hand, the $\psi$-variable is constructed
from the RFG model where $A\to\infty$ at constant density (and thus
contains no finite-mass dependences), although, as discussed in more
detail below, it does reflect some of
the missing-energy content in the problem (see also \cite{Cenni97}).

Thus, each approach has its own merits.
To bridge the gap at least partially, it is useful to shift the energy
$\omega$ to
\begin{equation}
\omega' \equiv \omega - E_{shift}
\label{omegapm}
\end{equation}
by an amount $E_{shift}$ to be chosen empirically (see the next section
where we discuss the choices made for the shift and  allow $E_{shift}$ 
to take on values other than $E_s$, the separation energy).  In the familiar
$y$-scaling analysis, already one usually does not use $E_s$ as would be
demanded if strictly adhering to the PWIA, but rather lets the shift
``float'' to allow the quasielastic peak to occur in the correct 
position. This usually results in a somewhat larger value for the shift
and probably reflects the fact that implicitly one is trying to build in
some aspects of the initial-state physics such as the average removal
energy --- the average of the separation energies of the various shells
making up the nuclear ground state --- but also some aspects of 
FSI which can also produce a shift in the position
of the quasielastic peak. Actually, the difference between the strict
interpretation as a separation energy and the empirical value that 
emerges is typically rather small.

We then adopt the same strategy when proceeding from the RFG 
starting point and  introduce dimensionless variables as above,
$\lambda' \equiv \omega'/2 m_N$ and $\tau' \equiv \kappa^2 -   {\lambda'}^2$,
so that in parallel with Eq.~(\ref{psi0}) we have
\begin{equation}
{\psi_0}' \equiv \psi_0 [\lambda \to \lambda'] 
  = \frac{2}{\eta_F} \biggl[ \sqrt{\lambda' (1+\lambda')} 
   -\kappa \biggr] 
  = \Upsilon_\infty ({\widetilde \lambda}=\lambda')
\label{psi0pm}
\end{equation}
and Eq.~(\ref{psiappx})
\begin{equation}
\psi' \equiv \psi [\lambda \to \lambda'] 
  = {\psi_0}' \biggl[ 1 + \sqrt{1+1/4\kappa^2} \frac{1}{2} \eta_F
  {\psi_0}' + {\cal O} [\eta_F^2] \biggr] .
\label{psipm}
\end{equation}
\begin{figure}
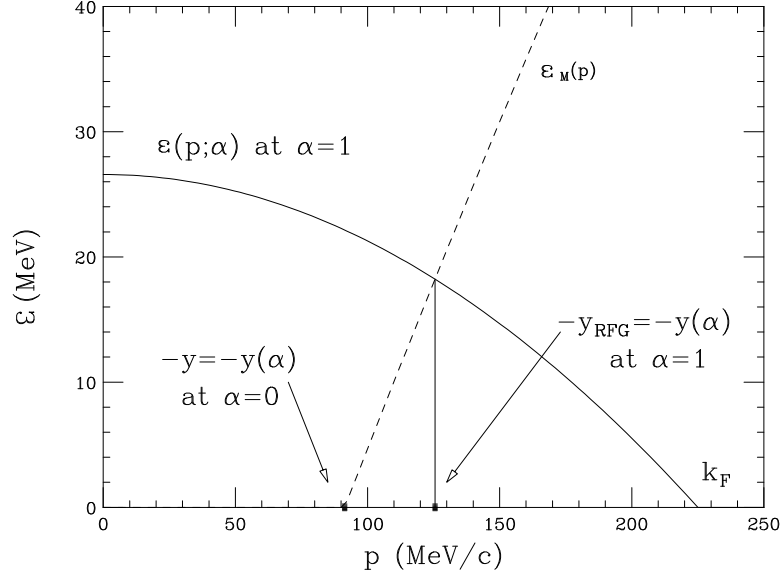

$$
\BoxedEPSF{fig1.pss scaled 600}  
$$
\caption{Missing-energy versus missing-momentum.
The various curves are described in the text; they are shown for
$k_F=$ 225 MeV/c, $E_{shift}=$ 20 MeV, $q=$ 700 MeV/c and $\omega=$ 200
MeV.}
\label{emissfig}
\end{figure}

A simple extension of the strict RFG model should help in providing an
understanding of the missing-energy content retained in defining a 
scaling variable. The RFG has a spectral function which is non-zero
along the line:
\begin{equation}
{\cal E}_{RFG}(p)=\sqrt{k_F^2+m_N^2}-\sqrt{p^2+m_N^2}
\label{emissRFG}
\end{equation}
that ``on the average'' \cite{Cenni97} incorporates the shell structure
of a typical (heavy) nucleus. Instead let us use
\begin{equation}
{\cal E}(p;\alpha)\equiv \alpha {\cal E}_{RFG}(p) ,
\label{emissAlpha}
\end{equation}
such that when $\alpha=1$ we recover the RFG model, but when $\alpha=0$
we have ${\cal E}=0$, the constraint used in defining the familiar
$y$ variable (see Eq.~(\ref{yusual}) in the Appendix). In Fig.~\ref{emissfig}
we show ${\cal E}(p;1)$ together with ${\cal E}_M(p)$ from 
Eqs.~(\ref{maxmissappx1},\ref{maxmissappx2}) at $q=$ 700 MeV/c and 
${\widetilde\omega}=$ 180 MeV 
--- namely for typical kinematics for the low-$\omega$ side of 
the quasielastic peak.
As in the RFG model (see \cite{Barbaro98}), the intersection of the two
curves occurs at the value of missing-momentum $p$ that defines the scaling
variable. For $\alpha=0$ this occurs at $-y$, the usual scaling variable
given approximately by $y_\infty$ in Eq.~(\ref{yinfty}) (for simplicity here
we have taken $\omega'={\widetilde\omega}$); for $\alpha=1$ it occurs at 
$-y_{RFG}$ given in Eq.~(\ref{yyrfg}) (but, of course, with $\omega$ shifted).
More generally one obtains something very similar to Eqs.~(\ref{psiappx}) and
(\ref{psipm}), namely
\begin{equation}
y(\alpha) = y_{\infty} \biggl[ 1 + \alpha \sqrt{1+1/4\kappa^2} \frac{1}{2} 
  \eta_F  {\psi_0}' + {\cal O} [\eta_F^2] \biggr] .
\label{yAlpha}
\end{equation}
Clearly what emerges is the following: the term in Eq.~(\ref{psipm})
containing a first-order correction to the ``minimal'' approximation to the
shifted RFG scaling variable, ${\psi_0}'$, is the one above involving
$\alpha$. When $\alpha=0$ (the usual $y$ definition) no missing-energy
dependence is taken into account, whereas with $\alpha\neq 0$ (as in the
RFG) some average dependence on ${\cal E}$ is incorporated.

Indeed, if circumstances warranted, it is straightforward to generalize
these ideas to devise still more scaling variables that build in the
best features of both the traditional PWIA-motivated extreme and the
RFG model extreme, or to go beyond in attempting to take into account
whatever we know about the missing-energy dependence of realistic
spectral functions. However, as the results given in the next section
show, such fine-tuning is apparently not needed at the present stage
of our understanding of superscaling.

Finally, having obtained dimensionless scaling variables $\psi$, $\psi'$
and $\Upsilon$ (together with approximations to them, as discussed above),
 we introduce a {\em dimensionless version of the scaling function} as 
suggested by the RFG model \cite{Alberico88,Barbaro98}
\begin{equation}
f \equiv k_F\times F .
\label{littlef}
\end{equation}
Not only does the RFG model contain scaling of the {\em first kind} so
that $f$ (or $F$) becomes independent of $q$ at high momentum transfers,
retaining dependence only on the scaling variable $\psi$, but it also
contains scaling of the {\em second kind} wherein $f$ is independent
of $k_F$ to leading order in $\eta_F^2$. What results for this model is
\begin{equation}
  f_{RFG}(\psi) = \frac{3}{4} (1 - \psi^2)\theta(1 - \psi^2) \left[ 1 + 
    (\eta_F\psi/2)^2 + \cdots \right] .
\label{fRFG}
\end{equation}
When both types of scaling
occur as they do for the RFG model we call the behavior {\em superscaling}.
In the next section we proceed to examine the degree to which scaling of 
the various kinds does or does not occur for measured unseparated
cross sections.

\section{ Scaling of Cross Sections: Results }
\label{sec:results1}
In this section we use the unseparated electron-nucleus inclusive 
scattering data presently
available to test the idea of superscaling for kinematics below the 
quasielastic peak and also try to add some insight into the various 
reasons which lead 
to the well-known fact that at large electron energy loss non-scaling 
behavior is observed. We restrict our attention to nuclei heavier than 
$^3$He, as the lightest nuclei are known to have momentum distributions that 
are very far from the ``universal'' one which is at the basis of the 
superscaling idea. 
\begin{figure}
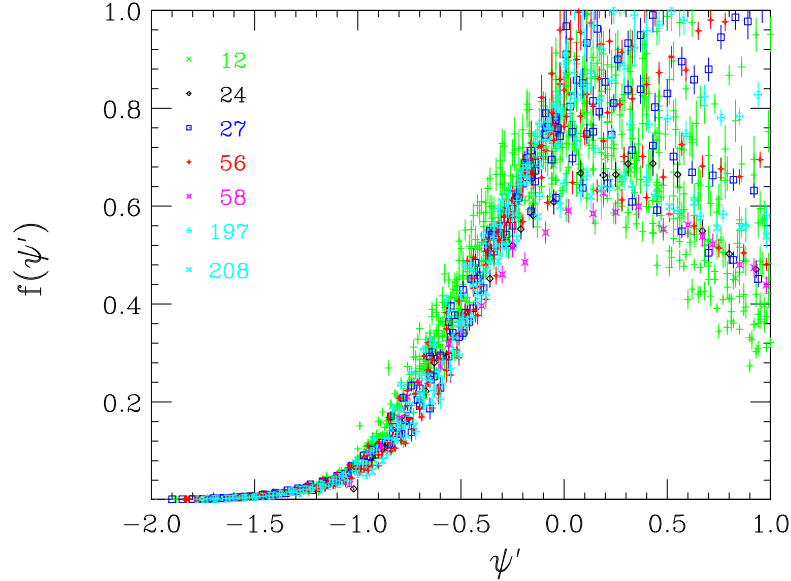

$$
\BoxedEPSF{fig2.pss scaled 600}  
$$
\caption{Scaling function $f(\psi ')$ as function of 
$\psi '$ for all nuclei $A \ge $ 12 and all kinematics. The values of $A$ 
corresponding to different symbols are shown in the insert.}
\label{di321}
\end{figure}

Data on inclusive electron-nucleus scattering for a series of nuclei 
($A =$ 4--208), but only one set of kinematics, were obtained early on 
by Whitney {\em et al.} \cite{Whitney74}. For helium, additional data at low
$q$ 
were measured by  Zghiche {\em et al.}, 
Dytman {\em et al.}, Meziani {\em et al.}, Sealock {\em et al.} 
 and von Reden {\em et al.} 
\cite{Whitney74,Zghiche94}-\nocite{Dytman88,Meziani92,Sealock89}\cite{Reden90};
high-$q$ data were obtained by Day {\em et al.} and  Rock {\em et al.}
 \cite{Day93,Rock82a}. 
For carbon, low momentum transfer data are available from  experiments 
performed by Barreau {\em et al.}, Baran {\em et al.} and O'Connell 
{\em et al.}  \cite{Barreau81}-\nocite{Barreau83,Baran88}\cite{Connell87};
 at high $q$ cross sections are available from the experiments of Day 
{\em et al.} and Heimlich {\em et al.} \cite{Day93,Heimlich74}. For 
oxygen an experiment has been performed by Anghinolfi {\em et al.} 
\cite{Anghinolfi96}. For 
medium-weight nuclei the data available include those for aluminum at
high $q$ measured by Day {\em et al.} \cite{Day93}, and the ones for calcium 
measured by Deady {\em et al.}, Meziani {\em et al.},  Yates {\em et al.} and 
Williamson {\em et al.} \cite{Deady86}-\nocite{Meziani85,Yates93}\cite{Williamson97} at 
low $q$. For 
iron experiments have been performed by  Altemus {\em et al.}, Meziani 
{\em et al.}, Baran {\em et al.}, Sealock {\em et al.} and Hotta {\em et al.}
 at low $q$ \cite{Altemus80,Meziani85,Baran88,Sealock89,Hotta84}; at high
$q$ measurements have been made by  Day {\em et al.} and Chen {\em et al.}
\cite{Day93,Chen91}. For heavy nuclei inclusive
cross sections have been measured by Day {\em et al.} for gold at high $q$
\cite{Day93},   
and by Zghiche {\em et al.}, Blatchley {\em et al.} and Sealock {\em et al.}
for nuclei between tungsten and uranium  at low $q$ 
\cite{Zghiche94,Blatchley86,Sealock89}.  

Not all of these data can be used, however, as some have not been corrected 
for radiative effects,  are known to have problems such as ``snout 
scattering'' or have a floating normalization;  some data are only 
available in the form of figures, but not as numerical values, and thus are 
not  useful in the present context. 

To begin with (see also \cite{Donnelly99}), we have taken the available data 
for the nuclei 
$A =$ 12...208 that meet our criteria for inclusion and have analyzed 
them in terms of scaling in the variable $\psi '$. Since $\psi '$ is 
defined in Eqs.~(\ref{psi0pm},\ref{psipm}) in terms of the Fermi momentum, 
appropriate values of $k_F$ had to 
be selected: specifically, we use 220, 230, 235 and 240 MeV/c for 
C, Al, Fe, Au, with intermediate values for the intermediate nuclei. The 
definition of the scaling variable also involves the choice of an 
appropriate ``shift'' energy (see Eq.~(\ref{omegapm})). This  energy accounts 
for the effects of
both the binding in the initial state and the interaction strength in the 
final state. In 
practice we use an energy that goes from 15 to 25 MeV for nuclei C...Au; the
results are quite insensitive to the exact choice.
\begin{figure}
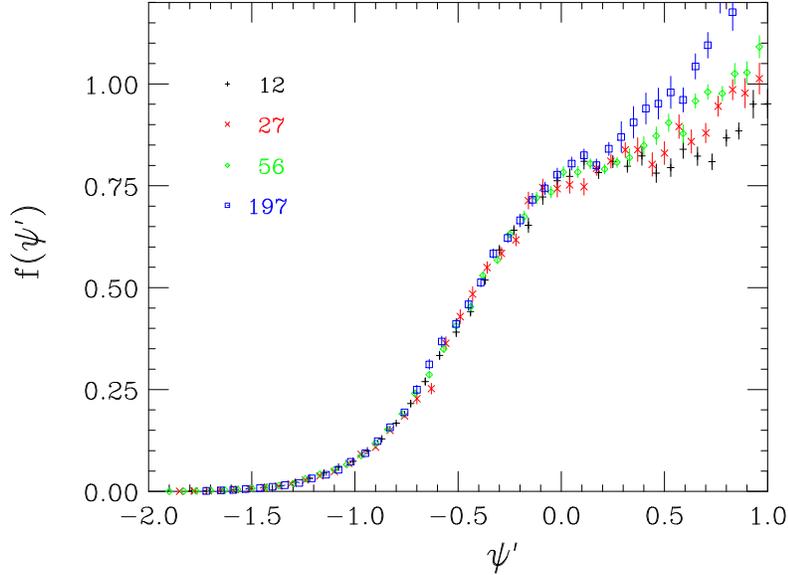

$$
\BoxedEPSF{fig3.pss scaled 600}  
$$
\caption{Scaling function for C, Al, Fe, Au and fixed 
kinematics ($q \approx$ 1000 MeV/c).}
\label{di421}
\end{figure}

Figure \ref{di321} shows the scaling function $f(\psi ')$ defined in
Eq.~(\ref{littlef}) for all kinematics
(energies, angles, momentum transfers) and all available nuclei meeting our
selection criteria. We observe reasonably
successful superscaling behavior for values of $\psi ' < 0$, while for 
$\psi ' >0$ the superscaling property is badly violated. The latter is to be 
expected, as there processes other than quasielastic scattering --- 
meson-exchange currents (MEC), $\Delta$-excitation, deep inelastic scattering 
--- contribute to the cross section, whereas the scaling as discussed in this 
paper only applies to quasielastic scattering.

In order to understand better the deviations from ideal scaling, below we
take different cuts through the data. The presently available data unfortunately
involve strong correlations in the kinematics employed: as the momentum 
transfer increases, the longitudinal (L) to transverse (T) cross section 
ratio for quasielastic scattering decreases. At the 
same time, the higher-$q$ data are taken at more forward angles. A separation of
the influence of the different driving factors such as $q$, L/T ratio and 
$A$-dependence is therefore not straightforward.

\begin{figure}
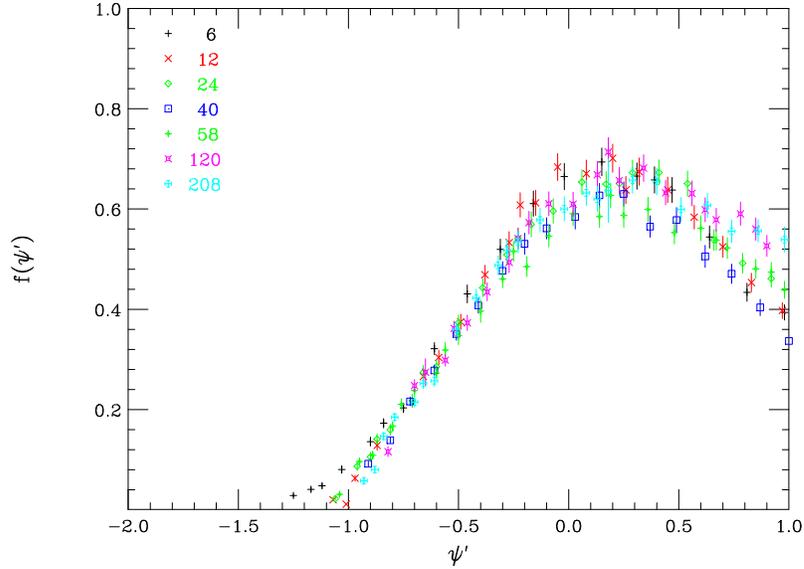

$$
\BoxedEPSF{fig4.pss scaled 600}  
$$
\caption{Scaling function for Li, C, Mg, Ca, Ni, Sn, Pb and fixed 
kinematics ($q \approx$ 460 MeV/c). }
\label{di311}
\end{figure}

In order to disentangle some of these less-than-perfect superscaling effects
at $\psi ' < 0$, we show in Fig.~\ref{di421}  the function $f(\psi ')$ for 
the series of nuclei $A=$ 12...197, but for fixed kinematics (electron energy 
3.6 GeV, scattering angle 16$^\circ$, where $q$ varies only mildly over
the range shown).
The quality of scaling of the second kind in the region $\psi ' <0$ is quite 
amazing, showing that insofar as the removal of the $A$-dependence is 
concerned the superscaling works extremely well and, importantly, that 
the deviations from superscaling observed in Fig.~\ref{di321} for 
$\psi ' < 0$ do {\em not} arise from the $A$-dependence. The scaling of the 
second kind works very well.

As similar quality of superscaling is found when analyzing other momentum 
transfers where a set of data for $A =$ 6...208 is available. As an example 
in Fig.~\ref{di311} we show the lower-$q$ data from the experiment of Whitney 
{\em et al.} \cite{Whitney74} taken at  500 MeV electron
energy and 60$^\circ$ scattering angle.

\begin{figure}
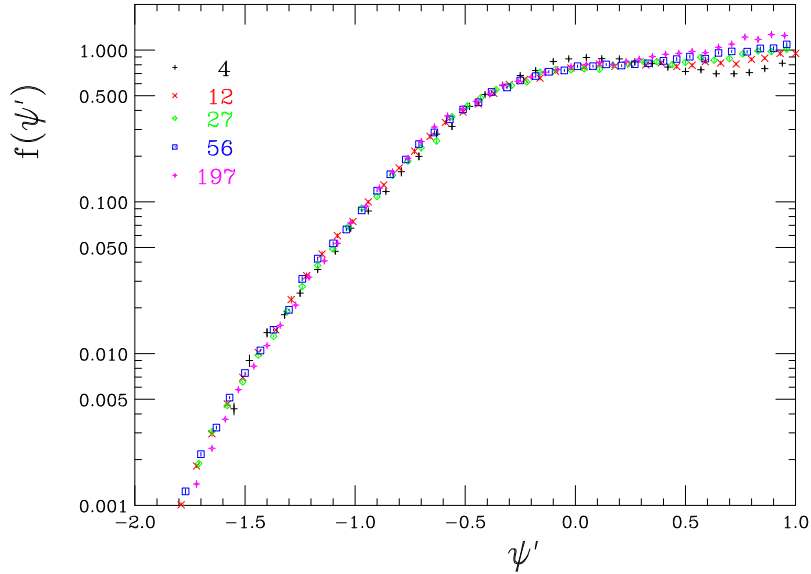

$$
\BoxedEPSF{fig5.pss scaled 600}  
$$
\caption{Scaling function for nuclei $A =$ 4--197 and fixed kinematics 
($q \approx$ 1000 MeV/c) on a logarithmic scale. }
\label{di423}
\end{figure}

Figure \ref{di423} shows the same data as those used in Fig.~\ref{di421} 
on a logarithmic scale, demonstrating that the superscaling property extends 
to large negative values of $\psi '$, values which in PWIA correspond to 
large momenta 
for the initial nucleon. A priori, this feature is not  predicted within the 
RFG model used to motivate the choice of $\psi '$. It can be understood, 
however, from the theoretical results for the momentum distribution of 
nuclear matter as a function of the nuclear matter density where, for 
different 
nuclear matter densities, the tail of  the momentum distribution is a 
near-universal function of $k/k_F$ \cite{Baldo90}. Since at large $k$ we 
deal with short-range properties of the nuclear wave function \cite{Benhar94b},
for finite nuclei and large momenta we can employ the Local Density  
Approximation (LDA), within which the nuclear momentum distribution (spectral 
function) is then a weighted  average over the corresponding nuclear matter 
distributions. This means that the large-momentum tail of the nuclear momentum 
distribution also scales with $k_F$, a dependence that is removed when using 
$\psi '$.  

In order to emphasize the quality of this superscaling in the tail, in  
Fig.~\ref{di423} we have also included the data for $^4$He which were taken 
under the same kinematical conditions as the other sets ($k_F=200$ MeV/c,
$E_{shift}=15$ MeV). While at 
$\psi '$=0  the superscaling function $f(\psi ')$ for $^4$He is about 15\% 
higher than for heavier nuclei, a consequence of the  sharper peak of the 
momentum distribution $n(k)$ at $k \approx 0$ for such a light nucleus, the 
scaling function for $^4$He agrees perfectly with the one for heavier nuclei 
when $\psi ' < - 0.3$. This reflects the fact that the tail of the momentum
distribution $n(k)$ at large $k$ is determined by the short-range properties
of the N--N interaction. 
  
Part of the $A$-dependent increase of $f(\psi ')$ at large $\psi '$ results 
from the increase of $k_F$ in proceeding from light to heavy nuclei. This 
amounts to an increase of the width of the 
quasielastic peak (i.e., before scaling with $k_F$) and a
correspondingly increased overlap with non-quasifree 
scattering processes ($\Delta$-excitation, $\pi$-production,...). 
At the same time, the increasing average density of 
the heavier nuclei also leads to an increase in contributions of two-body
processes such as MEC which are strongly density-dependent \cite{VanOrden81}. 
This, however, appears not to be the only cause for the rise (see the 
discussions in Sec.~\ref{sec:lt}).   
\begin{figure}
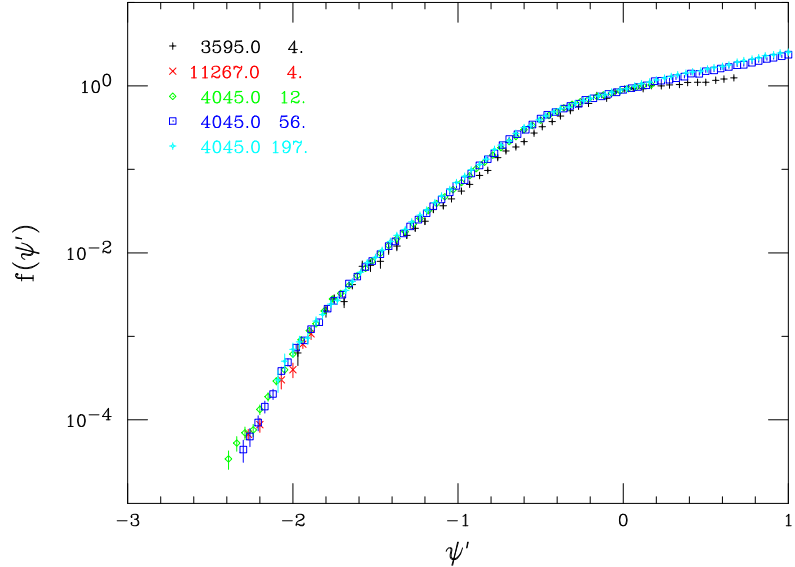

$$
\BoxedEPSF{fig6.pss scaled 600}  
$$
\caption{Scaling function  for nuclei $A =$ 4--197 at 
$q \approx$ 1650 MeV/c. The 4.405 GeV data have been taken at 23$^\circ$
scattering angle, the $^4$He data at 25$^\circ$ and 8$^\circ$.}
\label{di49}
\end{figure}

\begin{figure}
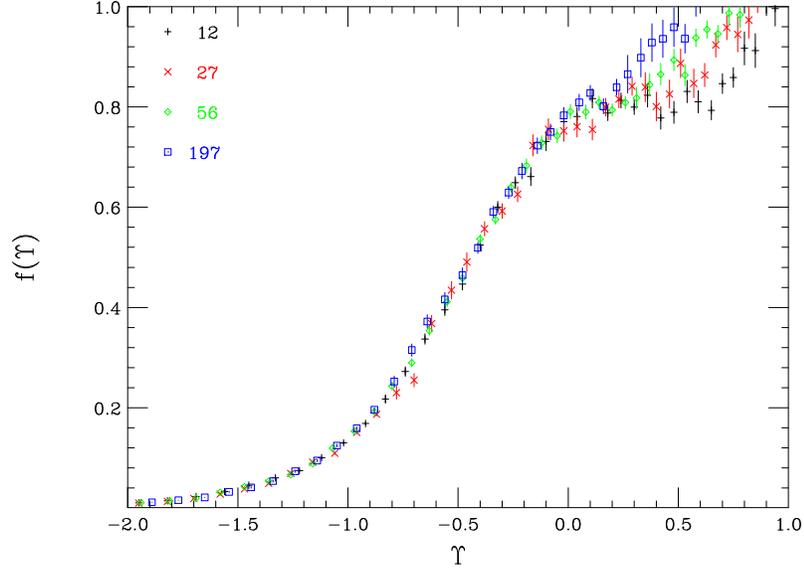

$$
\BoxedEPSF{fig7.pss scaled 600}  
$$
\caption{ Scaling function $f(\Upsilon)$ for nuclei $A =$ 12--197 
at 3.6 GeV energy and 16 degrees scattering angle as a function of
the scaling variable $\Upsilon$.}
\label{di424}
\end{figure}

Recently, the inclusive scattering data on C, Fe and Au have been extended 
to more negative values of $\psi'$ by an experiment performed at 
TJNAF \cite{Arrington99}. 
The higher product of beam current and spectrometer 
solid angle allowed Arrington {\em et al.} to measure cross sections a 
hundred times smaller
than previously accessible. In Fig.~\ref{di49} we show the scaling function 
for the set of data $A =$ 12--197 that extends to the most negative values 
of $\psi'$ reached, together with previous data on $A = 4$  \cite{Day93,Rock82a}
that also extend to rather large values of $|\psi '|$. 
Figure~\ref{di49} shows that the scaling of the second kind extends out to
the most negative values of the scaling variable presently accessible. 
At values of $\psi ' <$ --2 the scaling function seems to drop more 
rapidly with increasing $|\psi '|$, a feature that at present is not yet
understood; however, we continue to observe very high quality scaling of
the second kind.  
\begin{figure}
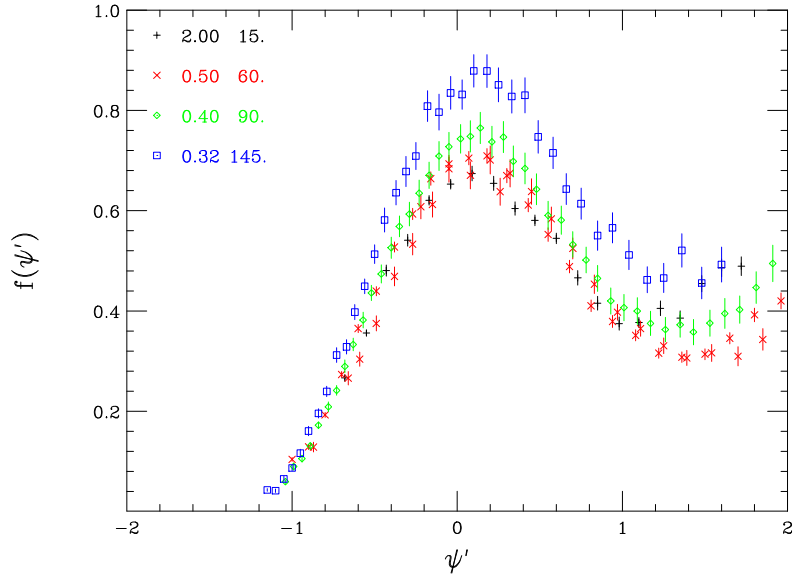

$$
\BoxedEPSF{fig8.pss scaled 600}  
$$
\caption{ Scaling function for $^{12}$C at 
approximately constant $q \approx$ 500 MeV/c, but
varying angle. The energies (in GeV) and angles (in degrees) of the different 
data  sets are identified in the plot.}
\label{di37}
\end{figure}

We have mentioned in the previous section that superscaling in terms of 
the variables $\psi '$ and $\Upsilon$ (see Eq.~(\ref{ups})) can be expected 
to be quite similar; data 
that scale in one variable can be expected to scale in the other one as well. 
As an example for this we show in Fig.~\ref{di424} the data of 
Fig.~\ref{di421} in terms of $f(\Upsilon)$.
In this paper we have concentrated on $\psi '$ rather than $\Upsilon$ since 
the former is more
directly related to the RFG model that motivated superscaling in the 
first place, and that allowed us to introduce $k_F$ as a physical scale 
used to define a dimensionless scaling variable. In the previous section we 
have given the relation between $\Upsilon$ and $\psi '$, pointing
out the difference in the treatment of the distribution in 
missing-energy which distinguishes the two scaling variables. The results
show that, at the large values of $q$ and $\omega$ of interest here, the scaling
behavior may be analyzed in terms of either variable.
\begin{figure}
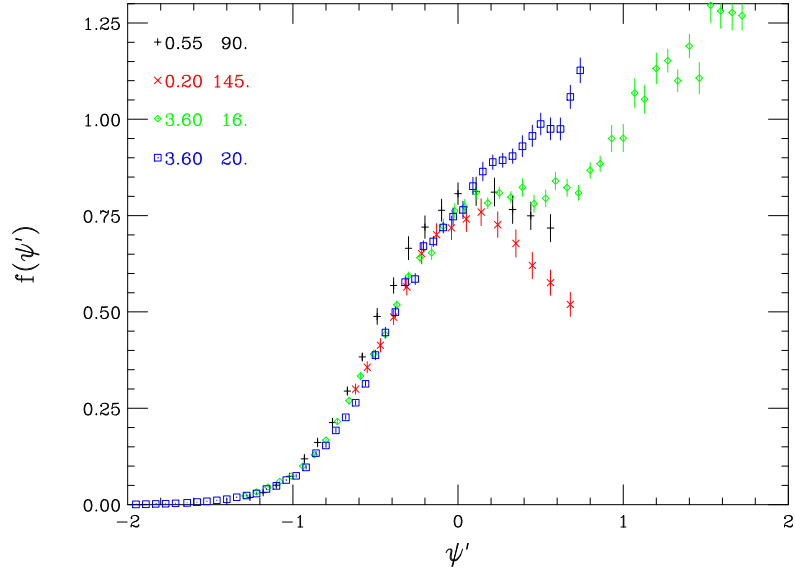

$$
\BoxedEPSF{fig9.pss scaled 600}  
$$
\caption{Scaling function for $^{12}$C and roughly constant ratio of the longitudinal
and transverse e-p elastic cross section, as a function of the momentum 
transfer~$q$.}
\label{di41}
\end{figure}

In order to locate the origin of non-scaling when all kinematics are
considered together (Fig.~\ref{di321}), for one nucleus ($^{12}$C) 
we have selected data sets corresponding to roughly constant momentum transfer 
$q$, but variable scattering angle (due to the discrete nature of the sets of 
data available, the choice of sets at ``constant $q$'' is only an approximate 
one). With increasing scattering angle, the 
ratio of the longitudinal to the total (longitudinal plus transverse) 
cross section decreases. For example, for e-p elastic scattering, which 
is characteristic of quasielastic electron-nucleus scattering,
and  for the momentum transfer $q \approx $ 500 MeV/c of Fig.~\ref{di37}, 
this ratio goes  from 0.5 at the highest energy and smallest scattering
angle down to less than 0.1 (i.e., the angle $\chi_{TL}$ defined in the next 
section, which characterizes the TL ratio, goes from about 45$^\circ$ down to 
about 15$^\circ$ for the four sets of kinematics chosen). The rise of 
$f(\psi ')$ 
for decreasing longitudinal contribution clearly shows that the dominant
piece responsible for non-scaling is the transverse one, as expected from the
dominantly transverse nature of $\Delta$-excitation and MEC. The violation of
scaling is still comparatively small as the momentum transfer of the data
in Fig.~\ref{di37} is small.

Figure \ref{di41} gives a different cut through the data presently available. 
Here we plot sets of data with an approximately constant longitudinal/transverse
ratio for the elastic e-p cross section, but varying momentum transfer $q$ 
(here $\chi_{TL}$ is roughly constant, typically within a few degrees of
25$^\circ$). The range of $q$ covered here is 320 to 1120 MeV/c. Clearly, the 
non-scaling contribution at $\psi ' > 0$ rises rapidly with
$q$. Part of this increase of the contribution at $\psi ' > 0$ originates from 
the fact that the quasielastic peak has a width that increases with $q$; as
a consequence, the overlap of the quasielastic response and the contribution
from other processes grows with increasing $q$. This effect is not immediately
obvious when looking at the representation in Fig.~\ref{di41} as the use of 
the scaling variable $\psi '$ removes this increase for the quasielastic
contribution. It arises partially because pion production (including via 
$\Delta$ production) moves to the left with increasing $q$ and so overlaps 
with the quasielastic peak (at $\psi'=0$). Other processes may also play
a role, in particular those involving MEC (to which we briefly return in
the next two sections) and eventually at high energies those involving deep 
inelastic electron-nucleon scattering. We note that,
according to Fig.~\ref{di37}, these non-quasifree contributions are basically 
of transverse nature. This suggests going further and attempting to
disentangle the L and T contributions to scaling --- we proceed to do so
in the next section. 

\section{ Scaling of Separated Responses \label{sec:lt}}

The various ways of looking at sub-samples of the data discussed above
show clearly that the violation 
of superscaling is basically the same as the one observed in conventional 
scaling applied to a single nucleus. The deviations from scaling
observed in Fig.~\ref{di321} can be understood in terms of the $q$- and 
L/T-dependence for a single nucleus. In particular, let us proceed to
extend the formalism of Sec.~\ref{sec:formal1} by
writing longitudinal and transverse versions of
Eq.~(\ref{Ftotal}). Starting from the cross section written in terms of
the individual response functions $R_L$ and $R_T$,
\begin{equation}
\frac{d^2 \sigma}{d \Omega_e d \omega} =
   \sigma_M [v_L R_L (\kappa, \lambda) + v_T R_T (\kappa, \lambda)] ,
\label{LTsep}
\end{equation}
we have from Eq.~(\ref{Ftotal}) that
\begin{equation}
F=\frac{v_L R_L + v_T R_T}{v_L G_L + v_T G_T} ,
\label{Ftotalt}
\end{equation}
with $G_{L,T}$ given in Eqs.~(\ref{G-RFG}).
The longitudinal and transverse analogs of this equation are
\begin{equation}
F_L = \frac{R_L}{G_L}, \qquad F_T = \frac{R_T}{G_T} .
\label{LTscaling}
\end{equation}
As we shall see below, it proves useful to study the difference between
these two quantities
\begin{equation}
\Delta F \equiv F_T-F_L.
\label{deltaf1}
\end{equation}
If the reaction mechanism in the quasielastic region is strictly
(quasifree) knockout of protons and neutrons, then one has
$F_L(\kappa,\psi)=F_T(\kappa,\psi)=F(\kappa,\psi)$,
namely, one has $\Delta F(\kappa,\psi)=0$. In light of the discussions
in the present work we might call this universality scaling of the 
{\em zeroth kind}.

The dimensionless analogs of Eq.~(\ref{littlef}) are given by
\begin{eqnarray}
 f_{L,T} &\equiv& k_F F_{L,T} \\
 \Delta f &\equiv& k_F \Delta F = f_T - f_L .
\label{scalfns}
\end{eqnarray}
The universality contained in the RFG model predicts that 
\begin{equation}
f_L=f_T=f, 
\label{scalingall}
\end{equation}
where the last is given in Eq.~(\ref{fRFG}), and moreover that
\begin{equation}
\int d\psi f^{RFG} (\psi) = 1 + \frac{1}{20} \eta_F^2 + \cdots ,
\label{CSR}
\end{equation}
which is closely related to the Coulomb sum rule \cite{Barbaro98}.

It is convenient to express the relationship amongst the $f$'s
(or the $F$'s) in the form
\begin{equation}
f \equiv \sin^2 \chi_{TL} f_L + \cos^2 \chi_{TL} f_T ,
\label{harmonic}
\end{equation}
where the angle $\chi_{TL}$ characterizes the TL content of the scattering
($\chi_{TL}=0^\circ \leftrightarrow $ all T; $\chi_{TL}=90^\circ 
\leftrightarrow$ all L). It is a solution to the equation
\begin{eqnarray}
\tan^2 \chi_{TL} &\equiv& (v_L/v_T)(G_L/G_T) \label{betaTLx} \\
&\cong& \frac{\left({\widetilde G}_E/{\widetilde G}_M \right)^2}
  {\tau +2\kappa^2 \tan^2 \theta_e/2} .
\label{betaTL}
\end{eqnarray}
We thus have a direct relationship for $\Delta f$ in terms of $f$ and 
$f_L$ (see below):
\begin{equation}
\Delta f = (f-f_L)/\cos^2 \chi_{TL} ,
\label{deltaf2}
\end{equation}
where this can be written in terms of
Eqs.~(\ref{betaTLx},\ref{betaTL}) 
using the fact that $1/\cos^2 \chi_{TL}=1+\tan^2 \chi_{TL}$.
In the data sets considered in the previous section the angle $\chi_{TL}$
varies considerably: specifically, in Figs.~\ref{di421}, \ref{di423} and 
\ref{di424} it is within a few degrees of 29$^\circ$; in Fig.~\ref{di311} 
within a few degrees of 43$^\circ$; in Fig.~\ref{di49} within a few 
degrees of 20$^\circ$; while for Figs.~\ref{di37} and \ref{di41} the values 
were stated earlier.

\begin{figure}
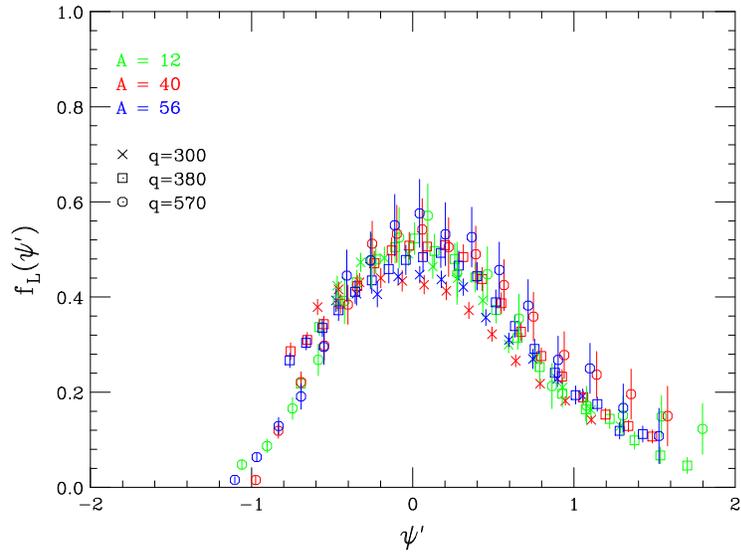

$$
\BoxedEPSF{fig10.pss scaled 600}  
$$
\caption{Scaling function \protect{$f_L(\psi ')$} from the longitudinal 
response.}
\label{fpsipl}
\end{figure}

\begin{figure}
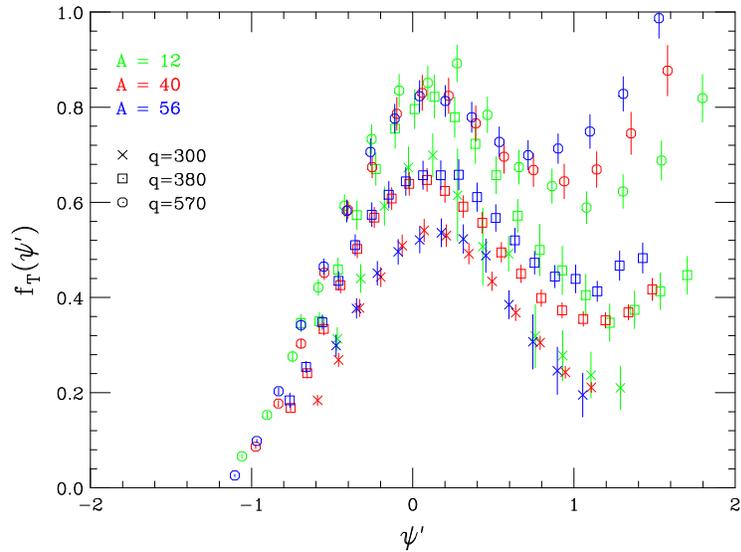

$$
\BoxedEPSF{fig11.pss scaled 600}  
$$
\caption{Scaling function \protect{$f_T(\psi ')$}
 from the transverse response.}
\label{fpsipt}
\end{figure}

Scaling of the first kind of the longitudinal and transverse response 
functions has been studied some time ago by Finn {\em et al.} \cite{Finn84}.
These authors found that, over the region $q$ from 250 to 550 MeV/c, the 
longitudinal response of $^{12}$C showed good scaling, while the transverse 
response did not. More recently \cite{Williamson97} $\psi'$-scaling of
L and T responses was 
investigated for the case of $^{40}$Ca.

As pointed out above, the longitudinal and transverse 
contributions to the cross sections should --- for quasielastic scattering 
in PWIA --- also show scaling of the second kind, in fact to the same 
response function; see Eq.~(\ref{scalingall}). In Figs.~\ref{fpsipl} and 
\ref{fpsipt}  we compare the scaling functions $f_L(\psi ')$ and 
$f_T(\psi ')$  obtained by Jourdan \cite{Jourdan96c,Jourdan98} who performed a 
longitudinal/transverse separation of the data for selected nuclei and
the lower momentum transfers ($<$580 MeV/c) where enough data for such
a separation are available. 
Within the error bars of the separated data the longitudinal response does 
scale to a universal curve, and as shown by Jourdan \cite{Jourdan96c}, the 
integral over this curve does fulfill the Coulomb sum rule (Eq.~(\ref{CSR})). 
Figures~ \ref{fpsipl} and \ref{fpsipt} also show that the basic problem
in quasielastic electron-nucleus scattering is the {\em excess in the 
transverse response} at large energy loss which grows rapidly with   $q$. 
It is {\em not} a lack of strength in the longitudinal response, as was 
claimed by some of the earlier determinations of the longitudinal response.

Figures \ref{fpsipl} and \ref{fpsipt} also explain the behavior of $f(\psi ')$ 
found in the previous figures at larger $q$'s. A transverse contribution is
clearly present which, up to $\psi ' \approx + 0.6$, has roughly the shape of 
the quasielastic peak. This leads to an excess of the transverse over the 
longitudinal strength without modifying the shape of the response in the 
region of the peak. For $\psi'< +0.6$ scaling of the {\em second kind} is
quite good, whereas scaling of the {\em first kind} is not. At larger 
$\psi '$ a (likely different) non-scaling contribution comes in at the 
larger $q$, which is much more important for the heavier nuclei; that is,
even scaling of the second kind is broken there.

\begin{figure}
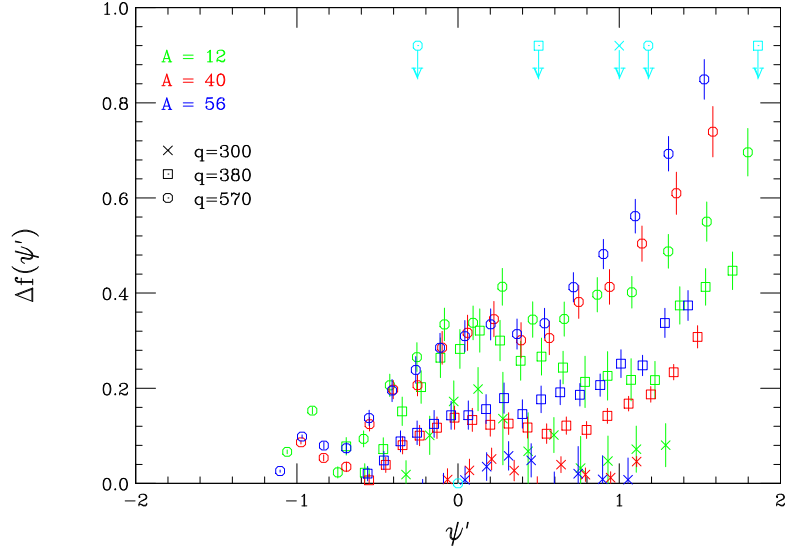

$$
\BoxedEPSF{fig12.pss scaled 600}  
$$
\caption{Difference between the transverse and longitudinal scaling 
functions, \protect{$\Delta f(\psi ')$}. 
The arrows  indicate the values of $\psi '$ for $\pi$-production on
the free nucleon and the nucleus at the three values of $q$.}
\label{fpsipts}
\end{figure}

In order to illustrate this point better, in Fig.~\ref{fpsipts} we show 
the {\em difference} between the transverse and longitudinal scaling
functions, $\Delta f(\psi ')$ defined in Eq.~(\ref{scalfns}).

As the longitudinal scaling functions for the different momentum transfers and 
mass numbers define an essentially universal curve (see Fig.~\ref{fpsipl}), we
have taken a bin-wise average of the data for the higher $q$'s of 
Fig.~\ref{fpsipl} in order to 
obtain the mean longitudinal response with smaller fluctuations. As the 
response at the lowest value of $q \approx $ 300 MeV/c is still subject to Pauli
blocking, we use only the data at the higher $q$'s to determine this 
universal longitudinal response. Indeed, to the extent that one believes
this to be {\em the} superscaled $f_L$, it is then possible to use the
unseparated function $f$ for {\em any} nucleus with $A\geq 4$ and {\em any}
(large enough) momentum transfer via Eq.~(\ref{deltaf2}) to determine 
$\Delta f$ and hence $f_T$. The difference in Fig.~\ref{fpsipts} shows 
that part of the excess transverse strength does indeed display a peak at the 
location of the maximum of the quasielastic response, $\psi '$=0. The strength
at larger $\psi '$, corresponding to larger electron energy  loss and of
presumably different origin, rises rapidly with increasing $q$.
  
Much of the strength of $\Delta f$ at $\psi ' < 0$ is below the threshold
for pion production on a nucleus with $A \geq 12$ (and even more so for 
quasifree production). This is shown by the arrows in Fig. \ref{fpsipts} 
which indicate, for the various $q$'s, the position of the $\pi$-production
threshold both on the nucleus and on the free nucleon --- we consider 
the latter
to be the more relevant one, since coherent production on the entire nucleus 
is expected to be very small. The presence of large excess transverse 
strength {\em below} $\pi$-threshold means that some other mechanism must be 
identified as its source. Various possibilities exist, for example, 
non-quasifree reactions in which the FSI are different for nucleon knockout 
via the L and T contributions of the electromagnetic current, cluster knockout 
and two-body MEC contributions; we return to touch upon some of these in the
next section.

 While the qualitative message  of Fig.~\ref{fpsipts} is clear,  we note that 
the numerical values of the difference in strength given there  should be 
treated with some care. It is clear that some processes playing a role
in accounting for the excess strength (for example, one-particle
emission via MEC) arise from {\em coherent} contributions to quasielastic 
scattering, and hence make any quantitative interpretation of 
$\Delta f$ less straightforward than would be the case when only
incoherent processes are present.

\section{Discussion and Conclusions} 
\label{sec:concl}
 We have analyzed the existing high-quality data on electron-nucleus 
quasielastic scattering for all nuclei $A$ = 4--238. We observe that, upon 
use of the proper scaling variable $\psi '$ (or, alternatively, 
$\Upsilon=y/k_F$), 
the data on the low-$\omega$ side of the quasielastic peak 
($\psi'$ or $\Upsilon  < 0$)  show {\em superscaling} behavior:  the scaling 
functions are not only independent of momentum transfer, but coincide for the 
different $A$  once the leading $k_F$ dependence is removed in the
manner discussed in this work. The former we call scaling of the {\em first
kind} and the latter scaling of the {\em second kind}. 

The main part of this work has been performed using the scaling variable 
$\psi '$ introduced within the context of the RFG model as motivation for the 
definition of a dimensionless scaling variable using the Fermi momentum as 
a scale. We have also discussed the relationship between $\psi '$ 
and  the usual scaling variable $y$, and shown that the two 
variables, which integrate somewhat different initial-state physics, yield
similar results. Indeed, the superscaling property is found in terms of 
both $\psi '$ and $\Upsilon$.

The $A$-independence of the  superscaling function actually 
is much better realized than the $q$-independence of 
the normal scaling; scaling of the first kind is known to be violated 
due to effects from FSI (mainly at very 
negative $\psi'$) and MEC, pion production and excitation 
of internal  degrees of freedom of the nucleon (mainly at $\psi ' > 0$). 
This observation of superscaling
allows us to conclude that, in the integral sense reflected through inclusive 
scattering, different nuclei have a more or less universal spectral function 
(momentum distribution) once the obvious dependence on the Fermi momentum
$k_F$ is 
removed. This universality is not restricted to the region of the 
quasielastic peak ($|\psi '| < 1$); the superscaling extends to larger 
values of $|\psi'|$ and hence to large values of the nucleon momentum in the
nucleus, a fact which may stem from the universal properties of nuclear 
spectral functions that arise from short-range N--N interactions 
insofar as they can lead to a scaling in terms of $k/k_F$.

Superscaling turns out to be particularly useful when dealing with the 
separated longitudinal and transverse responses. In quasielastic scattering 
for all large enough momentum transfers and all nuclear mass numbers $A\geq 4$
both of these responses should scale to the {\em same} function to which the 
unseparated data also scale. In particular, to the extent that the limited 
scope of the available data permits a test to be made, we find that the 
{\em longitudinal} response does scale to a universal curve, and that the
integral of the superscaled result satisfies the Coulomb sum rule.
  
However, when using superscaling to investigate further the reasons that lead 
to the observed non-scaling at electron energy loss 
$\omega' > |Q^2|/2m_N$ ($\psi ' >0$) for
individual nuclei, we find that the main problem resides in the 
{\em transverse} strength, which increases rapidly with increasing $q$, and 
less rapidly with increasing $A$. Some of the increase with $q$ is clearly 
related to the increasing overlap of the quasielastic contribution with
the $\Delta$-peak (which is predominantly transverse), the growing 
contribution of $\pi$ production and at the highest energies deep-inelastic 
scattering. This cannot provide all of the excess in $\Delta f$, however, 
since it clearly occurs below threshold for meson production, as well as at
higher energy loss.

Thus, the good quality of the scaling of the second kind is not entirely 
understood. Various sources for the excess transverse strength can be
identified, ranging from FSI effects to contributions from MEC\null. The former
could yield some T/L differences through spin-isospin many-body contributions
arising from RPA correlations or effects involving correlated knockout of
nucleon pairs (for instance, the $^1S_0 \to ^3S_1 + ^3D_1$ channel is
primarily transverse), although it is completely unclear what breaking of
scaling of the first or second kinds might be produced and whether the 
transverse/longitudinal excess could be so explained. Indeed, for example, 
one can argue that some contributions such as those stemming from 
short-range FSI are reasonably A-independent.
This is not the case for the contribution of MEC. For instance, the treatment
of Van Orden and Donnelly \cite{VanOrden81} shows that the 2p-2h MEC 
superscaled response contains an additional dependence of approximately 
$k_F^3$ and hence strongly breaks the second-kind scaling behavior. In fact,
those calculations yielded a rather small 2p-2h contribution --- which is
consistent with what is observed. Other studies
\cite{Dekker94,Amaro94} confirm this behavior. In particular, even 
calculations involving a dynamic
$\Delta$ propagator, such as those of Dekker et al. \cite{Dekker94}, while
providing somewhat larger 2p-2h MEC contributions, do not provide so much
that they disagree with the second kind scaling behavior (although note that
for reasons we do not yet understand 
recent work \cite{Gadiyak98} appears to be in conflict with the earlier
treatments). Furthermore, it should be pointed out that MEC effects enter
in the 1p-1h sector as well as in the 2p-2h sector. In
\cite{Alberico90} (and confirmed in \cite{Amaro94}) it was 
seen that the former interfere destructively with the one-body contributions
and therefore tend to lower the total 1p-1h transverse response --- when all
is added up the total MEC effect at and below the quasielastic peak is
found to be rather small. Clearly the reasons for the good quality of the 
scaling of the second kind and the limits that may be imposed on processes
such as MEC-mediated 2p-2h excitations certainly merit further theoretical
investigation.

In summary, superscaling, when applied to these separated responses, allows 
one in a 
particularly obvious way to make a point that recently has become increasingly 
clear: for inclusive electron-nucleus scattering the poorly understood
contribution is the {\em transverse} one, and not the longitudinal one as was 
usually claimed before the work of Jourdan \cite{Jourdan96c} in which 
reliable values for the longitudinal response were extracted.

\section*{Acknowledgements} 

The authors would like to thank J. Jourdan for providing the scaling functions
for the longitudinal and transverse responses and C.F. Williamson
for useful discussions during the course of this work. 

This work was supported in part by funds provided by the U.S. Department
of Energy under cooperative research agreement
\#DF-FC02-94ER40818, and by the Swiss National Science Foundation.  


\section*{Appendix }
\label{sec:appendix}

\subsection*{Kinematical Relationships}
Using Eq.~(\ref{scriptE}) it may be shown that
when ${\cal E}=0$ (its smallest value) the minimum and maximum values of 
the missing-momentum occur at $|y|$ and $Y$, respectively, where~\cite{Day90}
\begin{eqnarray}
  y &=& \frac{1}{2W^2} \biggl\{\left(M^0_A + \omega\right) 
    \sqrt{W^2-\left(M^0_{A-1} + m_N\right)^2} 
    \sqrt{W^2-\left(M^0_{A-1}-m_N\right)^2} \nonumber \\
    && \qquad - q\left[W^2+\left(M^0_{A-1}\right)^2 
    - m_N^2\right]\biggr\} \label{yusual} \\
  Y &=& \frac{1}{2W^2} \biggl\{\left(M^0_A + \omega\right) 
    \sqrt{W^2-\left(M^0_{A-1} + m_N\right)^2} 
    \sqrt{W^2-\left(M^0_{A-1}-m_N\right)^2} \nonumber \\
    && \qquad + q\left[W^2+\left(M^0_{A-1}\right)^2 - m_N^2\right]\biggr\}
\label{y-variables}
\end{eqnarray}
with as usual $W= \sqrt{\left(M^0_A + \omega\right)^2 - q^2}$. The variable 
$y=y(q,\omega)$ may be used together with $q$ to replace the pair of variables 
$(q,\omega)$ and is well-suited to quasielastic electron scattering, since 
the quasielastic peak occurs near $y=0$, with $y < 0$ corresponding to the 
so-called ``$y$-scaling region'' which is the focal point of this work, whereas 
$y > 0$ corresponds to the resonance region and beyond to deep-inelastic 
scattering. Expanding in inverse powers of the daughter mass one has
\begin{equation}
y=y_\infty \biggl[ 1 - \biggl( \frac{ \sqrt{m_N^2+(q+y_\infty)^2} }
  { q+y_\infty } \biggr)
  \frac{y_\infty}{2 M_{A-1}^0} +{\cal O} [ (M_{A-1}^0)^{-2} ] \biggr],
\label{yexpand}
\end{equation}
where $y_{\infty}$ is given in Eq.~(\ref{yinfty}).
The upper limit may similarly be
expanded for large $M_{A-1}^0$, yielding
\begin{equation}
Y\cong 2 q \biggl[ 1 - \frac{ \sqrt{m_N^2+(q+y_\infty)^2} }{M_{A-1}^0}
 \biggr] + y.
\label{Yappx}
\end{equation}

Another useful relationship needed in some of the discussions presented in 
Sec.~\ref{sec:formal1} is that for the maximum value of missing-energy allowed 
for given $(q,\omega)$ and given missing-momentum $p$. One finds that 
${\cal E}$, which is essentially the missing-energy minus the separation 
energy $E_s$, has as its maximum value
\begin{eqnarray}
{\cal E}_M (q,y;p) &= \sqrt{m_N^2+(q+y)^2}-\sqrt{m_N^2+(q-p)^2}
   \nonumber \\
   & \qquad\qquad +\sqrt{(M_{A-1}^0)^2+y^2}-\sqrt{(M_{A-1}^0)^2+p^2} 
   \nonumber 
\end{eqnarray}
\begin{eqnarray}
 \stackrel{M_{A-1}^0 \to \infty}\longrightarrow 
    &\sqrt{m_N^2+(q+y)^2}-\sqrt{m_N^2+(q-p)^2} - 
    (p^2-y^2)/2M_{A-1}^0 \nonumber \\
 \stackrel{q \to \infty}\longrightarrow &(p+y) -
    \Bigl( \sqrt{(M_{A-1}^0)^2+p^2}-\sqrt{(M_{A-1}^0)^2+y^2} \Bigr) 
     \nonumber \\
 & \qquad \stackrel{M_{A-1}^0 \to \infty}\longrightarrow (p+y) - 
    (p^2-y^2)/2M_{A-1}^0 .
\label{maxmiss}
\end{eqnarray}
In the main part of the paper we employ only the $M_{A-1}\to\infty$ limit
as in Eqs.~(\ref{maxmissappx1},\ref{maxmissappx2}).

All of the kinematic relationships given above do not depend
on the choice of dynamical model beyond the assumption of nucleon
knockout. 

\subsection*{PWIA and the cc1 Off-shell Prescription}

If following common practice one invokes the PWIA for the reaction, then a 
nucleon of energy
\begin{equation}
E (p,{\cal E}) = M_A^0 -\sqrt{(M_{A-1}^0)^2 + p^2} -{\cal E}
\label{Etrans}
\end{equation}
and momentum $p$ is struck by the virtual photon and is ejected from the
nucleus as a plane-wave (on-shell) with energy $E_N$ and momentum $p_N$.
The kinematics of the reaction require the struck nucleon to be
off-shell; that is, $E\neq {\bar E}$, where ${\bar E}\equiv (m_N^2+p^2)^{1/2}$.
In fact the off-shellness may be characterized by the quantity
\begin{eqnarray}
  \rho (p,{\cal E}) &\equiv& \frac{ {\bar E}-E}{2 m_N} \nonumber \\
  &=& \frac{1}{2 m_N} \Bigl[ \Bigl( \sqrt{m_N^2+p^2}-m_N \Bigr) +
   \Bigl( \sqrt{(M_{A-1}^0)^2 + p^2}-M_{A-1}^0 \Bigr) \nonumber \\
   && \qquad\qquad +{\cal E} + E_S \Bigr] \geq E_S/2 m_N. 
\label{offshellness}
\end{eqnarray}
In PWIA the cross section is given as the product of 
the half-off-shell single-nucleon cross section and the nuclear spectral
function ${\widetilde S} (p,E)$ which gives the probability that a 
nucleon of 
momentum $p$ and energy $E$ is found in the nuclear ground state. We may then
write ${\widetilde S}$ as a function of $(p,{\cal E})$. 

For the single-nucleon cross section it is common practice to use
the cc1 prescription of De Forest~\cite{Forest83}. Then, integrating over 
azimuthal angles, summing over particles while assuming that the spectral 
function does not differ for protons and neutrons, and including the 
kinematic factor $E_N/q$ with $E_N=(({\bf q} + {\bf p})^2+m_N^2)^{1/2}$ 
(see~\cite{Day90}), one obtains the following for the single-nucleon
cross section:
\begin{equation}
  {\widetilde\sigma}_{eN}(q,\omega;p,{\cal E}) \equiv
   \sigma_M \Bigl[ v_L {\widetilde w}_L + v_T {\widetilde w}_T \Bigr],
\label{sigmatil}
\end{equation}
with $\sigma_M$ the Mott cross section and $v_{L,T}$ the usual
Rosenbluth kinematical factors, where the longitudinal (L) and transverse 
(T) cc1 contributions may be written:
\begin{eqnarray}
{\widetilde w}_L (q,\omega;p,{\cal E}) &=& \frac{1}{2\kappa\sqrt{1+\eta^2}}
  \Bigl(\frac{\kappa^2}{ {\bar\tau} }\Bigr) \Bigl[ {\widetilde G}_E^2 
  + \delta^2 ({\widetilde W}_2 + \Delta {\widetilde W}_1) \nonumber \\
  && \qquad\qquad +(1+{\bar\tau}) \Delta {\widetilde W}_1
     +(1+\tau) \Delta {\widetilde W}_2 \Bigr] \nonumber \\
{\widetilde w}_T (q,\omega;p,{\cal E}) &=& \frac{1}{2\kappa\sqrt{1+\eta^2}}
  \Bigl[ 2 {\bar\tau} {\widetilde G}_M^2 + \delta^2 
  ({\widetilde W}_2 + \Delta {\widetilde W}_1) \Bigr] .
\label{cc1}
\end{eqnarray}
Here we employ dimensionless variables $\kappa\equiv q/2m_N$, 
$\lambda\equiv \omega/2m_N$, $\tau\equiv\kappa^2-\lambda^2>0$, where
$\omega=E_N-E$. The cc1 prescription introduces the energy ${\bar E}$
given above and hence the ``equivalent on-shell energy transfer''
${\bar\omega}=E_N-{\bar E}$, with ${\bar\lambda}\equiv {\bar\omega}/2m_N$
and ${\bar\tau}\equiv \kappa^2-{\bar\lambda}^2$. We have also defined
$\eta\equiv p/m_N$, where then ${\bar E}/m_N=(1+\eta^2)^{1/2}$, and used the 
fact that:
\begin{eqnarray}
\delta^2 &\equiv& \frac{ {\bar\tau} }{\kappa^2} 
\Bigl( \frac{E_N+{\bar E}}{2 m_N} \Bigr)^2 - (1+{\bar\tau}) \\
&=& \frac{ {\bar\tau} }{\kappa^2} \Bigl[ 2{\bar\lambda}
(\sqrt{1+\eta^2}-1)+\eta^2 \Bigr] - \Bigl( \frac{{\bar\lambda}
-{\bar\tau}}{\kappa} \Bigr)^2,
\label{delta}
\end{eqnarray}
where the relationship 
$(E_N+{\bar E})/2 m_N= {\bar\lambda} + \sqrt{1+\eta^2}$ has been used
to obtain the result in Eq.~(\ref{delta}). The terms containing
$\delta^2$ as a factor enter because the struck nucleon is moving and
contribute whether or not the nucleon is off-shell. As discussed in the main
text, $\eta$ is typically small; therefore the first term in
Eq.~(\ref{delta}) is very small, being of order $\eta^2$. For the
second term in this equation we can use as an estimate
Eq.~(\ref{upsinfty}) and find that its contribution is also very
small, being of order $\eta_F^2$. Thus the terms in Eqs.~(\ref{cc1})
containing the factor $\delta^2$ are all seen to be very small.

The single-nucleon form factors enter Eqs.~(\ref{cc1}) in the following
combinations: 
\begin{eqnarray}
  {\widetilde G}_E^2 (\tau) &\equiv& Z G_{Ep}^2 + N G_{En}^2 \nonumber \\
  {\widetilde G}_M^2 (\tau) &\equiv& Z G_{Mp}^2 + N G_{Mn}^2 \nonumber \\
  \Delta {\widetilde G} (\tau) &\equiv& Z G_{Ep} G_{Mp} 
    + N G_{En} G_{Mn} 
\label{formfac}
\end{eqnarray}
where $G_{Ep,n}$ and $G_{Mp,n}$ are the familiar Sachs form factors and
are functions only of $\tau$, and then:
\begin{eqnarray}
  {\widetilde W}_1 (\tau) &\equiv& \tau {\widetilde G}_M^2 \nonumber \\
  {\widetilde W}_2 (\tau) &\equiv& \frac{1}{1 + \tau} 
    [ {\widetilde G}_E^2 +\tau {\widetilde G}_M^2 ] \nonumber \\
  \Delta {\widetilde W}_1 (\tau, {\bar\tau}) &=& 
    \frac{{\bar\tau}-\tau}{(1+\tau)^2}
    \left[ {\widetilde G}_E^2 + {\widetilde G}_M^2 
    -2 \Delta {\widetilde G} \right] \nonumber \\
  \Delta {\widetilde W}_2 (\tau, {\bar\tau}) &=& 
    \frac{{\bar\tau}-\tau}{(1+\tau)^2}
    \left[ {\widetilde G}_E^2 - {\widetilde G}_M^2 \right] .
\label{responses}
\end{eqnarray}
The form given here for the cc1 prescription is different from the
usual one~\cite{Forest83}, having been rearranged to bring out the strong 
resemblance to the on-shell form discussed below. Note that, when a nucleon
is moving but on-shell, since $\tau={\bar\tau}$ the last two responses 
are zero, $\Delta{\widetilde W}_{1,2}=0$. That also
implies, as expected, that no terms of the form $G_E G_M$ coming from 
$\Delta{\widetilde G}$ in Eq.~(\ref{responses}) can occur when on-shell,
although they do for the cc1 off-shell prescription. Finally, note
that these off-shell effects are all proportional to ${\bar\tau}-\tau$
which may be written
\begin{equation}
{\bar\tau}-\tau = \rho ( 2\lambda -\rho )
\label{offf}
\end{equation}
using Eq.~(\ref{offshellness}).

\subsection*{Scaling in the RFG Model}

We end this Appendix by collecting some of the exact expressions involved
in studies of the RFG model (see also \cite{Alberico88,Barbaro98}). First
the $\psi$-scaling variable is fully given by
\begin{equation}
  \psi = \frac{1}{\sqrt{\xi_F}} \frac{\lambda - \tau}
   {\sqrt{(1+\lambda)\tau + \kappa \sqrt{\tau (1+\tau)}}} ,
\label{psifull}
\end {equation}
where $\xi_F = \sqrt{1 + \eta_F^2} - 1$ and $\eta_F =
k_F/m_N$ are the dimensionless Fermi kinetic energy and momentum, respectively.
Approximations to this quantity were employed in the main part of the paper
for simplicity (see Eqs.~(\ref{psi},\ref{psiappx})), although computations 
were done with the full expression.

The exact RFG analog of Eq.~(\ref{Fscaling}) is:
\begin{equation}
F (\kappa, \psi) \equiv \frac{d^2 \sigma/d \Omega_e d \omega}
   {\sigma_M [v_L G_L (\kappa, \lambda) + v_T G_T (\kappa, \lambda)]} ,
\label{Ftotalfull}
\end{equation}
where we have made use of 
the usual lepton kinematical factors $v_L$ and $v_T$ and on-shell
single-nucleon responses $G_L$ and $G_T$ (see \cite{Barbaro98},
and also \cite{Amaro96,Jeschonnek98}) 
\begin{eqnarray}
G_L (\kappa,\lambda) &=& \frac{ (\kappa^2/\tau) 
    [ {\widetilde G}_E^2 + {\widetilde W}_2 \Delta ] }
    {2\kappa [1+\xi_F (1+\psi^2)/2]} \nonumber \\
  &=& \frac{\kappa}{2\tau} {\widetilde G}_E^2 + {\cal O}[\eta_F^2] \nonumber \\
G_T (\kappa,\lambda) &=& \frac{ 2\tau {\widetilde G}_M^2 
    + {\widetilde W}_2 \Delta }
    {2\kappa [1+\xi_F (1+\psi^2)/2]}  \nonumber \\
 &=& \frac{\tau}{\kappa} {\widetilde G}_M^2 + {\cal O}[\eta_F^2] .
\label{G-RFG}
\end{eqnarray}
Here~\cite{Alberico88} 
\begin{eqnarray}
  \Delta &=& \xi_F (1-\psi^2) \left[ \frac{\sqrt{\tau (1+\tau)}}{\kappa}
   + \frac{1}{3} \xi_F (1-\psi^2) \frac{\tau}{\kappa^2} \right] \nonumber \\
  &=& \frac{1}{2} (1-\psi^2) \eta_F^2 +{\cal O}[\eta_F^3].
\label{Delta}
\end{eqnarray}
The above approximations yield the expressions used in the main part of
the paper (see Eq.~(\ref{Ftotal})). Note that the on-shell 
limits of Eqs.~(\ref{cc1}) immediately give the behavior seen in
Eqs.~(\ref{G-RFG}), as they should.

\end{document}